%% file: main.tex
\documentclass[journal]{new-aiaa}
\usepackage[utf8]{inputenc}
\usepackage{textcomp}

\usepackage{graphicx}
\usepackage{amsmath}
\usepackage[version=4]{mhchem}
\usepackage{siunitx}
\usepackage{longtable,tabularx}
\usepackage{soul}
\usepackage{multirow}
\usepackage{booktabs} 
\usepackage{siunitx}  
\usepackage{todonotes}

\title{LES of Droplet Impingement: Application to Clean and Laser-Scanned Ice Shapes}

\author{Federico Zabaleta \footnote{Postdoctoral Fellow, Department of Mechanical Engineering, Center for Turbulence Research; fzabaleta@stanford.edu. AIAA Member}}
\affil{Center for Turbulence Research, Stanford University, Stanford, CA 94305}
\author{Brett Bornhoft\footnote{Cadence Design Systems, San Jose, CA, AIAA Member.}}
\affil{Cadence Design Systems, San Jose, CA 95134}
\author{Suhas S. Jain\footnote{Assistant Professor, George W. Woodruff School of Mechanical Engineering, AIAA Young Professional Member.}}
\affil{Flow Physics and Computational Science Lab, Georgia Institute of Technology, Atlanta, GA 30332}
\author{Sanjeeb T. Bose\footnote{Cadence Design Systems, San Jose, CA, AIAA Member.}}
\affil{Cadence Design Systems, San Jose, CA 95134}
\author{Parviz Moin\footnote{Franklin P. and Caroline M. Johnson Professor, Department of Mechanical Engineering, Center for Turbulence Research. Fellow AIAA. \\ A portion of this work was presented as Paper 2024-4162 at the AIAA AVIATION FORUM AND ASCEND 2024, Las Vegas, Nevada, July 29--August 2, 2024.}}
\affil{Center for Turbulence Research, Stanford University, Stanford, CA 94305}

\begin{document}

\maketitle

\begin{abstract}
The prediction of aircraft icing is conventionally performed using multishot simulation frameworks that fail to predict the progressive roughening of the ice surface. To understand roughness formation, we investigate droplet impingement on clean and laser-scanned rough ice shapes using a high-fidelity computational framework based on wall-modeled large-eddy simulations and Lagrangian particle tracking. This methodology is validated against experimental data for a NACA 23012 airfoil and a NACA 64A008 swept tail, accurately predicting collection efficiency and supercooled large droplet splashing. The framework is subsequently applied to laser-scanned rime ice geometries to quantify the impact of surface roughness on local impingement distributions. The results reveal that physical roughness induces a highly nonuniform collection efficiency, with droplet impingement intensely concentrated on upstream-faces of roughness elements, creating sheltered shadow zones immediately downstream. While the spanwise-averaged collection efficiency remains remarkably similar to that of an equivalent smooth body, idealized smooth surfaces completely suppress these localized impingement peaks. Ice accretion simulations demonstrate that this localized impingement creates a self-reinforcing feedback loop, actively amplifying existing roughness features over time. These findings provide a direct physical explanation for the formation of characteristic rime ice structures and highlight the critical role of local surface topology in the accretion process.

\end{abstract}

\section*{Nomenclature}

{\renewcommand\arraystretch{1.0}
\noindent\begin{longtable*}{@{}l @{\quad=\quad} l@{}}
$\overline{(\cdot)}$ & Filtered variable [-] \\
$\tilde{(\cdot)}$ & Favre (density-weighted) filtered variable [-] \\
$A_{bf}$ & boundary face area [m$^2$] \\
$c$    & airfoil chord length [m] \\
$C_p$ & pressure coefficient [-] \\
$d_p$ & droplet diameter [m] \\
$dA_s$ & infinitesimal area on the impact surface [m$^2$] \\
$d_{\text{imp}}$ & impinging droplet diameter [m] \\
$d_{\text{splash}}$ & secondary splashed droplet diameter [m] \\
$E$ & total energy per unit mass [J/kg] \\
$f_i$ & interaction force vector between particle and fluid [N] \\
$K$ & Mundo parameter [-] \\
$K_L$ & splashing parameter [-] \\
$k_s$ & equivalent sand-grain roughness height [m] \\
$k_{\max}$ & maximum peak-to-trough roughness height [m] \\
$k_{\text{rms}}$ & root-mean-square roughness height [m] \\
$m_{\text{imp}}$ & impinging droplet mass [kg] \\
$m_{\text{splash}}$ & splashed droplet mass [kg] \\
$m_p$ & particle mass [kg] \\
$\dot{m}_s$ & droplet mass flux at the body surface [kg/(m$^2\cdot$s)] \\
$\dot{m}_\infty$ & freestream mass flux of droplets [kg/(m$^2\cdot$s)] \\
$N_{\text{cv}}$ & number of control volumes [-] \\
$N_s$ & number of particles impacting $dA_s$ [-] \\
$Oh$ & Ohnesorge number [-] \\
$O_R$ & statistical overloading ratio [-] \\
$p$ & static pressure [Pa] \\
$Q_i^{sgs}$ & subgrid heat flux [W/m$^2$] \\
$Re_c$ & chord Reynolds number [-] \\
$Re_p$ & particle Reynolds number [-] \\
$s$ & streamwise coordinate [m] \\
$T$ & static temperature [K] \\
$t$ & time [s] \\
$t_{sim}$ & simulation averaging time interval [s] \\
$U_\infty$ & freestream velocity magnitude [m/s] \\
$u_i$ & fluid velocity vector components [m/s] \\
$u_{p,i}$ & particle velocity vector components [m/s] \\
$u_{n}$ & normal velocity component [m/s] \\
$u_{t}$ & tangential velocity component [m/s] \\
$x_i$ & cartesian coordinate vector components [m] \\
$x_{p,i}$ & particle position [m] \\
$\alpha$ & droplet impact angle [rad] \\
$\beta$ & collection efficiency [-] \\
$\Delta_{\min}$ & minimum cell dimension [m] \\
$\Lambda$ & parcel impingement rate [s$^{-1}$] \\
$\lambda$ & thermal conductivity [W/(m$\cdot$K)] \\
$\mu$ & dynamic viscosity [Pa$\cdot$s] \\
$\Phi$ & splashed mass fraction [-] \\
$\rho$ & fluid density [kg/m$^3$] \\
$\rho_p$ & particle density [kg/m$^3$] \\
$\tau_{ij}$ & viscous stress tensor [Pa] \\
$\tau^{sgs}_{ij}$ & subgrid stress tensor [Pa]
\end{longtable*}}
\setcounter{table}{0} 

\section{Introduction}
\lettrine{T}{he} impingement of supercooled liquid droplets on aircraft surfaces is the initial step in the formation of in-flight ice, a phenomenon that poses a significant and well-documented threat to aviation safety. Ice accretion alters the aerodynamic profile of wings and control surfaces, leading to performance degradation through increased drag, reduced lift, and a higher stall speed \citep{gentAircraftIcing2000}. The historical impact of this hazard in the United States is substantial; a review of the years 1982--2000 linked in-flight icing to 583 accidents and more than 800 fatalities \citep{pettyStatisticalReviewAviation2004}, and it continues to be a factor in modern aviation incidents \citep{nationaltransportationsafetyboardNTSBAviationAccidents2023}. Depending on the ambient conditions, ice can form with two distinct morphologies: rime and glaze ice. Rime ice occurs when droplets freeze completely upon impact, a process typically associated with low ambient temperatures, low liquid water content (LWC), and small droplet sizes. Because the droplets solidify immediately, the resulting ice shape is determined predominantly by the droplet impingement distribution, making its accurate prediction essential for correctly modeling the growth of rime ice. In contrast, glaze ice forms when only a fraction of a droplet's mass freezes upon contact, while the remainder forms a liquid film on the surface. This liquid water can then run back along the airfoil, freezing at a different location or being shed entirely, often leading to the formation of complex structures such as horns. Consequently, rather than depending solely on the initial impingement distribution, the growth of glaze ice is strongly governed by the underlying heat transfer processes  \citep{zabaletaLargeeddySimulationsConjugate2025}.

The prediction of aircraft icing relies on a numerical framework that leverages the large separation of timescales between the slow ice growth process and the much faster aerodynamic and droplet dynamics. Representing the ice growth as a series of quasi-steady steps allows a multistep/shot approach to be employed. First, the aerodynamic flow field around the geometry is computed (often using Reynolds-averaged Navier--Stokes closures \cite{moulaIcingSimulationResults2023,bellostaLagrangianEulerianAlgorithms2023}). The resulting velocity field is then used to solve a droplet transport model, which yields the local collection efficiency,  quantifying the rate of particle impingement on the surface. This collection efficiency then serves as an input to a thermodynamic model, where mass and energy conservation equations are solved over the airfoil/ice surface \citep{messingerEquilibriumTemperatureUnheated1953,myersExtensionMessingerModel2001} to determine the local ice growth rate. For a multishot simulation, this rate is used to evolve the surface geometry, and the entire sequence is repeated for each subsequent time step/shot.

Droplet dynamics are described through either  an Eulerian--Lagrangian approach \citep{shadStokesdependentDropletCollection2025,moulaIcingSimulationResults2023,trontinDescriptionAssessmentNew2017} or an Eulerian--Eulerian approach \citep{araujolimadasilvaAdvancingIceAccretionFoamSolver2025,radenacWorkflowPredictorCorrector2023}. The Lagrangian method tracks the trajectory of individual particles or parcels, facilitating the straightforward incorporation of particle--wall interaction effects, such as splashing. Alternatively, the Eulerian--Eulerian description models the dispersed phase as a continuous medium, offering continuity of the solution and often achieving lower computational costs for large three-dimensional configurations, albeit with difficulties in accurately accounting for secondary splashing droplets and shadow regions \citep{guardoneAircraftIcingModeling2025}.

While this quasi-steady, multishot framework can effectively predict the mean profile of rime ice accretions, as demonstrated in the 1st AIAA Ice Prediction Workshop (IPW 1) \citep{laurendeauSummary1stAIAA2022}, a significant shortcoming is its failure to predict the small-scale surface roughness characteristic of real ice. This idealization to a smooth surface represents a significant limitation, as roughness is responsible for an increase in skin friction, heat transfer, and early transition to turbulence. To bridge this gap, ice accretion codes use additional models to account for the effects of roughness, typically by approximating the nonhomogeneous surface with an equivalent sand grain roughness, $k_s$. This parameter is derived from empirical models, which can be based on either experimental correlations \citep{ignatowiczSurfaceRoughnessRANS2023} or simplified bead dynamics that represent the freezing of individual droplets \citep{ozcerFENSAPICENumericalPrediction2011}.

The aerodynamic significance of these small-scale features has been the subject of recent investigation. \citet{bellostaAssessingRelevantRoughness2024} demonstrated that the aerodynamic degradation associated with rime ice shapes is caused primarily by roughness. Specifically, when they analyzed a progressively smoothed version of a laser-scanned rime ice shape, they found that removing the smallest roughness scales significantly altered the aerodynamic behavior, with performance metrics approaching those of a clean airfoil. Furthermore, the underlying flow physics diverge significantly between smooth and rough shapes. \citet{bornhoftUseArtificialIce2025} noted that simulations on smooth ice shapes can introduce flow features like laminar separation, which are absent on real, rough surfaces where the boundary layer immediately transitions to turbulence. Therefore, a comprehensive understanding of how roughness develops and evolves is fundamental for improving the fidelity of aerodynamic degradation predictions.

Despite this recognized importance, standard multishot simulation frameworks fail to predict this evolution because the spatial and temporal averaging inherent in their methodology tends to produce artificially smooth updated ice geometries \citep{freschiTwoDimensionalMultiStepStochastic2025,zabaletaMultishotSimulationsRime2025}. By continuously evaluating the droplet impingement distribution over these progressively smoothed surfaces, multishot codes suppress the initial, highly localized impingement variability required to initiate discrete roughness elements. While empirical roughness models attempt to account for the resulting macroscopic aerodynamic and thermodynamic effects, the direct physical interaction between the incoming droplets and the highly irregular surface topology is completely omitted during the ice accretion process. This omission leaves the effect of small-scale surface irregularities on local droplet impingement largely unexplored, neglecting some of the physical mechanisms that govern the accretion process.

To explore the interactions of particles with small roughness features, the present work evaluates droplet impingement directly on laser-scanned rough surfaces. Furthermore, high-fidelity, scale-resolving simulations are used to provide an accurate and detailed description of the highly localized flow structures and transient particle trajectories surrounding these small-scale features. This study has two primary objectives. First, we validate a high-fidelity framework based on wall-modeled large-eddy simulations (WMLES) and Lagrangian particle tracking to predict collection efficiency in geometries with varying levels of complexity. Second, we investigate the effect of realistic ice roughness on the collection efficiency distribution, providing insight into the physical mechanisms that drive roughness development and growth. The remainder of this paper is organized as follows. The numerical formulation, including the governing equations and the splashing model, is described in Sec.~\ref{sec:NumMethods}. The validation of the methodology against experimental data for the clean NACA 23012 airfoil and the NACA 64A008 swept tail is presented in Sec.~\ref{sec:Validation}. In Sec.~\ref{sec:Roughness}, the model is applied to two laser-scanned rime ice geometries with different accretion exposure times to quantify the impact of surface roughness on local impingement distributions. The self-reinforcing nature of ice roughness growth is then analyzed in Sec.~\ref{sec:roughness_evolution}. Finally, the conclusions are summarized in Sec.~\ref{sec:conclusions}.

\section{Numerical methods}\label{sec:NumMethods}
The numerical simulations are conducted using the computational fluid dynamics code charLES, a low-dissipation, finite-volume solver \citep{bresLargeeddySimulationsCoannular2018}. This work solves the low-pass filtered, compressible Navier--Stokes equations for mass, momentum, and total energy,
\begin{equation}
    \frac{\partial \overline{\rho}}{\partial t} + \frac{\partial \overline{\rho}\tilde{u_i}}{\partial x_i} = 0, 
\end{equation}
\begin{equation}
    \frac{\partial \overline{\rho}\tilde{u_i}}{\partial t} + \frac{\partial \overline{\rho}\tilde{u_i}\tilde{u_i}}{\partial x_i} = - \frac{\partial \overline{p}}{\partial x_i} + \frac{\partial \tilde{\tau}_{ij}}{\partial x_i} - \frac{\partial \tilde{\tau}^{sgs}_{ij}}{\partial x_i},
\end{equation}
\begin{equation}
    \frac{\partial \overline{E}}{\partial t} + \frac{\partial \tilde{u_i} \overline{E}}{\partial x_i} + \frac{\partial \tilde{u_i}\overline{p}}  {\partial x_i} = \frac{\partial \tilde{\tau}_{ij}\tilde{u_i}}{\partial x_i} - \frac{\partial \tilde{\tau}^{sgs}_{ij}\tilde{u_i}}{\partial x_i} + \frac{\partial}{\partial x_i}\left( \lambda \frac{\partial \overline{T}}{\partial x_i}\right) - \frac{\partial \tilde{Q}^{sgs}_{i}}{\partial x_i}.
\end{equation}
The subgrid terms are closed using the dynamic Smagorinsky model \citep{germanoDynamicSubgridScale1991,moinDynamicSubgridScale1991}, and pressure, density, and temperature are related by the ideal gas law. 

Time integration is achieved using a five-stage explicit Runge--Kutta scheme, while spatial discretization is performed with second-order accurate schemes using skew-symmetric operators that conserve kinetic energy in the inviscid, zero Mach number limit, and also approximately preserve entropy in the inviscid, adiabatic limit \citep{honeinHigherEntropyConservation2004,chandrashekarKineticEnergyPreserving2013}. Given the high Reynolds number, an algebraic wall model is employed to calculate wall shear stress, enabling accurate results while maintaining a reasonable computational cost \citep{lehmkuhlLargeeddySimulationPractical2018}.

\subsection{Lagrangian-particle model}

The dynamics of the dispersed phase (droplets) are described using a one-way coupled Lagrangian particle approach. This approach is suitable for aircraft icing simulations because of the low volumetric concentrations of droplets, $\mathcal{O}(10^{-6}$), allowing droplet effects on the airflow and droplet--droplet interactions to be neglected \citep{Elghobashi1994}.
The trajectory of each individual particle is computed by integrating Newton's second law,
\begin{equation}
    \frac{d x_{p,i}}{d t} = u_{p,i},
\end{equation}
\begin{equation}
    m_p\frac{d u_{p,i}}{d t} = -f_i.
\end{equation}
Due to the high density ratio $(\rho_p/\rho\approx 1000)$ and the high disparity between freestream velocity and settling velocity [$\mathcal{O}(10^{-4})$], the momentum transfer is dominated by the drag force \citep{guardoneAircraftIcingModeling2025}. The interaction force is modeled using the drag correlation presented by \citet{schillerDragCoefficientCorrelation1935},
\begin{equation}
    f_i = 3\pi \mu d_p \left(1+0.15Re_p^{0.687}\right) \left( u_{p,i} - u_i \right).
\end{equation}

\subsection{Collection efficiency and splashing model}

The parameter that describes the impingement rate of particles over the geometry is the collection efficiency, $\beta$. It is defined as the ratio of droplet mass flux at the body surface ($\dot{m}_s$) to the freestream mass flux ($\dot{m}_\infty$). The collection efficiency, in a Lagrangian framework, is calculated as
\begin{equation}
    \beta = \frac{\dot{m}_s}{\dot{m}_\infty},\quad  \dot{m}_s= \lim_{t\to \infty} \dfrac{1}{dA_s t}\sum_{k=1}^{N_s(t)} m_p^k, \quad \dot{m}_\infty= U_\infty\text{LWC},
    \label{eq:betaEq}
\end{equation}
The triangle-trajectory intersection algorithm developed by \cite{mollerFastMinimumStorage2005} is employed to detect collisions between the particle and the surface mesh.

Droplet--wall interactions are critical in supercooled large-droplet (SLD) icing conditions. Following impact, droplets exhibit one of three behaviors: adhesion to the surface, rebounding, or splashing into multiple secondary droplets. When rebounding or splashing occurs, water mass is returned to the airflow, directly affecting the net collection rate. The likelihood and extent of splashing depend on droplet material properties (density, viscosity, surface tension), kinematic conditions (size, impact velocity), and geometric factors (impact angle). Generally, the tendency to splash is enhanced by larger droplets, higher impact velocities, shallower impact angles, and reduced impact frequencies.

The splashing model presented by \citet{wrightFurtherRefinementLEWICE2006}, widely used in ice accretion prediction \citep{wrightComparisonLEWICEGlennICE2008a,bellostaLagrangianEulerianAlgorithms2023}, is employed here. This framework treats rebounding as a limiting case of splashing where all impacting mass is ejected as a single secondary droplet with no surface adhesion. Splashing onset is governed by the dimensionless parameter $K_L$, defined as
\begin{equation}
    K_L = \frac{0.859\sqrt{K}(\rho/\text{LWC})^{0.125}}{\sin (\alpha)^{1.25}},
\end{equation}
where $K$ is the Mundo parameter \citep{mundoDropletwallCollisionsExperimental1995}, $\alpha$ is the impact angle, and $K$ relates to the Ohnesorge number ($Oh$) and particle Reynolds number ($Re_p$) via 
\begin{equation}
    K = Oh Re_p^{1.25}.
\end{equation}
For $K_L \leq 200$, complete adhesion occurs with no splashing. When $K_L > 200$, the splashed mass fraction ($\Phi$) is
\begin{equation}
    \Phi = \frac{m_{\text{splash}}}{m_{\text{imp}}} = \begin{cases}
        0 & K_L\leq200,\\
        0.7(1-\sin \alpha) \left( 1 - e^{-0.0092026(K_L-200)}\right) & K_L>200.
    \end{cases}
    \label{eq:ratiosplashing}
\end{equation}
Secondary droplet properties (diameter and velocity components) follow from
\begin{equation}
    \frac{d_{\text{splash}}}{d_{\text{imp}}} = 8.72 e^{-0.0281K},
    \label{eq:veldiamsplash}
\end{equation}
\begin{equation}
    u_{n,\text{splash}} = u_{n,\text{imp}} (0.3 - 0.002 \alpha ),
\end{equation}
\begin{equation}
    u_{t,\text{splash}} = u_{t,\text{imp}} (1.075 - 0.0025 \alpha ).
\end{equation}
At each droplet--surface collision, $K_L$ and $\Phi$ are computed to determine the outcome. The adhering mass fraction is accumulated on the impacted computational face, while the splashed fraction generates secondary droplets with properties given by Eq.~(\ref{eq:veldiamsplash}). When splashing produces multiple secondary droplets, a parcel representation \citep{drew1998theory} is used to maintain computational efficiency.

\section{Validation and best practices}\label{sec:Validation}
\subsection{NACA 23012 airfoil}\label{subsec:NACA23012}

The first validation case replicates the experiments conducted by Papadakis et al. \cite{papadakisWaterImpingementExperiments2004} on a clean NACA 23012 airfoil. The airfoil has a chord length $c = 0.9144$ m (36 in.) and is subjected to a freestream velocity of $78.23$ m/s ($Re_c = 4.57\times 10^6$) at an angle of attack of 2.5$^{\circ}$. From the experimental campaign, three droplet clouds with mean volume diameters (MVD) of 20 $\upmu$m, 52 $\upmu$m, and 111 $\upmu$m are simulated. For each MVD, the 27-bin droplet distribution provided in the experiment is reproduced in Table \ref{table:binSizes}. For the MVD~$=20$ $\upmu$m spray, all but two bins fall below the supercooled large droplet (SLD) threshold, usually defined as $d_p \geq 40 \upmu$m \cite{AirplaneEngineCertification2014}. In contrast, for the 52 $\upmu$m and 111 $\upmu$m distributions, 17 and 23 bins respectively are categorized as SLD, indicating that splashing effects are significant. Consequently, the MVD$=20$ $\upmu$m case is used here to validate the base methodology and demonstrate the convergence properties of the single-distribution approach, while the larger distributions are employed to assess the splashing model.

\input{table1}

\subsubsection{Computational Setup} \label{sec:CompSetup}

\begin{figure}[htbp]
    \centering
    \includegraphics[width=0.7\textwidth]{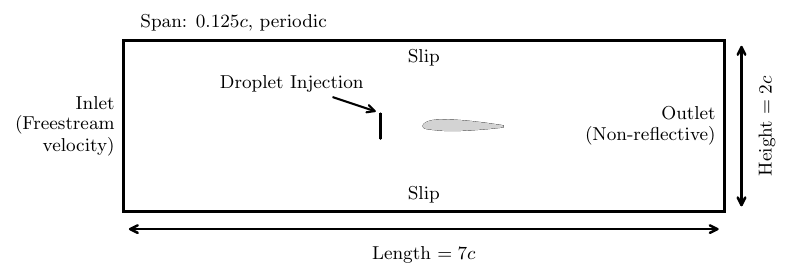}
    \caption{Schematic representation of the computational domain.}
    \label{fig:domain_schematic}
\end{figure}

As illustrated in Figure \ref{fig:domain_schematic}, the computational domain has a length of 7 times the chord length, a height of $2c$, and a span of $0.125c$. At the inlet, the mean freestream velocity was specified, and for the outlet a non-reflective boundary condition was used. Slip boundary conditions were used at the top and bottom of the domain, and periodic boundary conditions were used in the spanwise boundaries.

To evaluate grid convergence, the computational domain is discretized using a series of three unstructured Voronoi grids designated as coarse, medium, and fine. The total mesh size ranges from approximately $1.4$ million control volumes for the coarse grid to $11.2$ million control volumes for the fine case, with the intermediate level containing $\approx 3.0$ million control volumes. Successive grids are generated through homothetic refinement, effectively doubling the near-wall resolution at each step. The grid resolution, characterized by $c/\Delta_{\min}$, spans from approximately 1024 to 4096 across the three grids. This maximum level of refinement is concentrated within the first quarter of the chord length to resolve the strong flow acceleration near the leading edge before transitioning to coarser resolutions in the remaining portion of the airfoil.

Droplets are injected into the computational domain at a plane located $0.5c$ upstream of the airfoil's leading edge. This rectangular injection area spans the full width of the computational domain with a height of $0.3c$. Individual particles are seeded at random locations within this plane and are injected with an initial velocity equal to the freestream velocity. To minimize computational cost and restrict the dispersed phase to the regions of interest, the trajectory of each particle is tracked only until it either intersects the airfoil surface or passes the trailing edge, at which point it is eliminated from the domain.

\subsubsection{Time convergence of $\beta$}\label{sec:betaConvergence}

In typical atmospheric icing conditions, the low physical concentration of droplets requires simulating long physical times to obtain statistically converged collection efficiency distributions. This is further exacerbated when a continuous polydispersed cloud is modeled by injecting a statistical distribution of droplet sizes simultaneously \cite{zabaletaLargeEddySimulationSupercooled2024}. In a typical cloud, smaller droplets possess a high number density, while larger droplets are relatively rare. Despite their low number density, these large droplets have significantly higher mass per droplet, exerting a large influence on the local collection efficiency. Because these large droplets impact the surface much less frequently than the small droplets, obtaining a statistically converged impingement profile for the larger bins defines the total simulation time.

To overcome this limitation, the present methodology simulates each droplet bin individually \cite{zabaletaLargeEddySimulationSupercooled2024}. By decoupling the droplet sizes, the injection rate can be tailored specifically for each bin, ensuring that the low-frequency droplets achieve statistical convergence at the same rate as the smaller, high-frequency droplets. The individual $\beta$ distributions are then combined using their respective liquid water content (LWC) weighting factors to reconstruct the total collection efficiency of the polydispersed cloud.

To further reduce the computational cost of simulating long time domains, we employ a statistical overloading technique \cite{shadStokesdependentDropletCollection2025}. Because the droplet dynamics are one-way coupled, the number of simulated particles can be artificially increased to accelerate statistical convergence without altering the underlying physics or the carrier fluid flow. In the present simulations, the injection rate is calculated such that approximately 10 million particles are present simultaneously within the computational domain. This particle count was selected to provide a balance between statistical convergence rate and computational cost.

To estimate the statistical convergence rate of $\beta$ and mathematically justify the required simulation time, the impingement of discrete numerical droplets can be modeled as a spatial Poisson point process. Because droplet impingement in the Lagrangian framework occurs as a series of independent discrete events, the expected number of parcels impacting a given boundary face of area $A_{bf}$ over a simulation time $t$ is $E[N_t] = \Lambda  t$, where $\Lambda$ is the parcel impingement rate. This rate is a function of the local collection efficiency $\beta$, the freestream variables, the particle mass $m_p$, and the statistical overloading ratio $O_R$:
\begin{equation}
    \Lambda = \frac{\beta \, \text{LWC} \, U_\infty A_{bf}}{m_p} \times O_R
\end{equation}
For a Poisson point process, the relative statistical error, expressed as the coefficient of variation ($CV = \sigma/\mu$), scales inversely with the square root of the expected number of impacts:
\begin{equation}
    CV = \frac{1}{\sqrt{E[N_t]}} = \frac{1}{\sqrt{\Lambda t}}
\end{equation}

\begin{figure}
    \centering
    \includegraphics{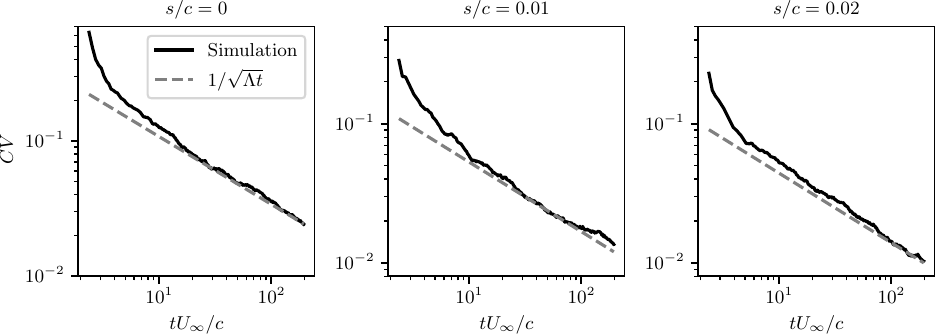}
    \caption{Time convergence of local collection efficiency ($\beta$) at three streamwise locations, compared to the $CV$ evolution of a Poisson point process.}
    \label{fig:time_convergence}
\end{figure}

To verify this theoretical convergence rate within the simulation framework, the spanwise variance of $\beta$ was tracked over time and compared with the theoretical prediction. Because the aerodynamic flow over the extruded geometry is statistically two-dimensional, any variation in $\beta$ along the span at a fixed chordwise location is attributable solely to Lagrangian statistical noise. Figure \ref{fig:time_convergence} compares the spanwise $CV$ at three streamwise locations ($s/c = 0$, $0.01$, and $0.02$) obtained from the simulations with the theoretical prediction. As illustrated, the simulation $CV$ decays proportionally to $t^{-1/2}$, falling directly on top of the analytical prediction curve. This confirms that the selected particle injection rate and statistical overloading ratio provide a predictable path to statistical convergence.

While this theoretical model validates the numerical implementation, it also highlights two fundamental computational challenges inherent to the Eulerian--Lagrangian framework. First, the $t^{-1/2}$ convergence rate dictates that reducing the statistical noise by half requires quadrupling the simulation time. Consequently, obtaining perfectly smooth $\beta$ distributions (i.e., driving $CV$ to near zero) through time integration alone becomes computationally very expensive. Second, the impingement rate is directly proportional to the face area ($A_{bf} \sim \Delta x^2$). Therefore, refining the computational mesh decreases the number of expected droplet impacts per cell. To maintain a constant level of statistical error on a finer mesh without increasing the physical simulation time, the overloading ratio $O_R$ must be increased inversely to $\Delta x^2$, which significantly increases the total number of tracked particles and the associated computational cost.

\subsubsection{Mesh convergence}

\begin{figure}
    \centering
    \includegraphics{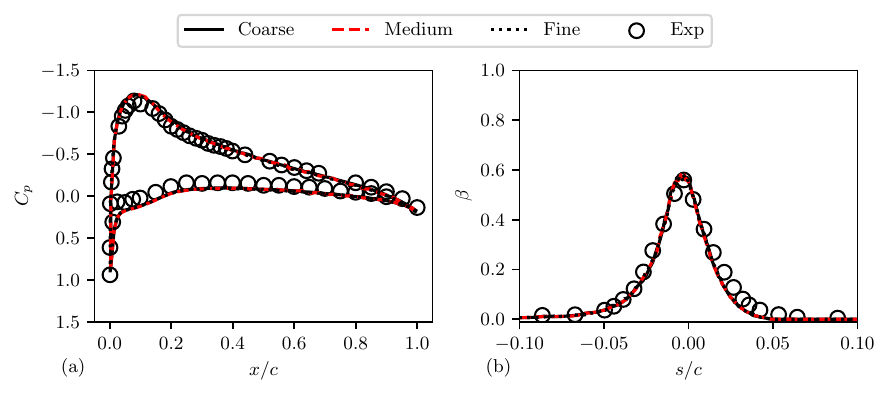}
    \caption{Grid convergence study showing the spanwise-averaged (a) pressure coefficient $C_p$ and (b) collection efficiency $\beta$ for the coarse, medium, and fine meshes, superimposed with experimental data \cite{papadakisWaterImpingementExperiments2004}.}
    \label{fig:mesh_convergence}
\end{figure}

Figure \ref{fig:mesh_convergence} presents the results of the grid refinement study, comparing the spanwise-averaged pressure coefficient (Fig.~\ref{fig:mesh_convergence}-a) and collection efficiency (Fig.~\ref{fig:mesh_convergence}-b) distributions across the coarse, medium, and fine grids. As observed in the $C_p$ profiles, the aerodynamic flow field is well resolved, with all three grid resolutions yielding virtually identical pressure distributions that closely match the experimental data. Similarly, the local collection efficiency distributions exhibit no significant variation among the tested meshes, accurately capturing the impingement peak and limits observed in the experiments. These results demonstrate that grid independence has been achieved, confirming that the coarse resolution is sufficient to accurately capture both the aerodynamic and droplet impingement physics without requiring further refinement. Given the low computational cost of these simulations, the medium grid resolution is utilized for the rest of this article.

\subsubsection{Number of droplet bins}

\begin{figure}
    \centering    \includegraphics{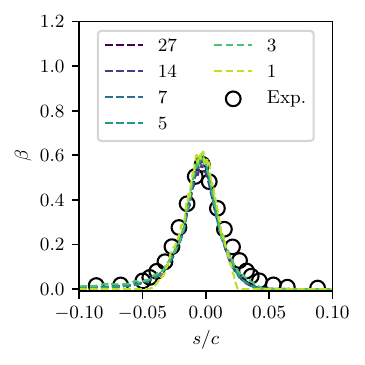}
    \caption{Sensitivity of $\beta$ to the number of simulated droplet bins for the 20 $\upmu$m MVD cloud.}
    \label{fig:Nbins}
\end{figure}

To evaluate the sensitivity of the collection efficiency to the discretization of the droplet size distribution, $\beta$ was calculated using 1, 3, 5, 7, 14, and 27 bins. When using fewer than 27 bins, the original distribution detailed in Table \ref{table:binSizes} is resampled into a reduced set of representative bins. This is achieved by lumping adjacent intervals, conserving the total liquid water content, and using an equivalent median volume diameter for each new bin. The resulting $\beta$ profiles corresponding to each discretization level are presented in Figure \ref{fig:Nbins}.

The profile obtained using only a single bin (the cloud MVD) accurately captures the peak impingement magnitude but significantly underpredicts the spatial extent of the impingement limits. Droplets larger than the MVD possess greater inertia and deviate less from their freestream trajectories as they travel around the airfoil. Consequently, these larger droplets impact further downstream along the surface, increasing the impingement limits. Representing the entire cloud with a single median diameter effectively ignores the effect of these high-inertia particles, leading to artificially narrow tails. The simulations utilizing between 3 and 27 bins exhibit virtually identical impingement limits and only marginal variations in the peak magnitude. This highlights that highly accurate impingement profiles can be achieved with significantly reduced bin samples, reducing the computational cost of the simulations. For the remainder of this article, seven bins are used to represent the droplet clouds.

\subsubsection{Collection efficiency for SLD distributions}

\begin{figure}
    \centering    
    \includegraphics[width=\textwidth]{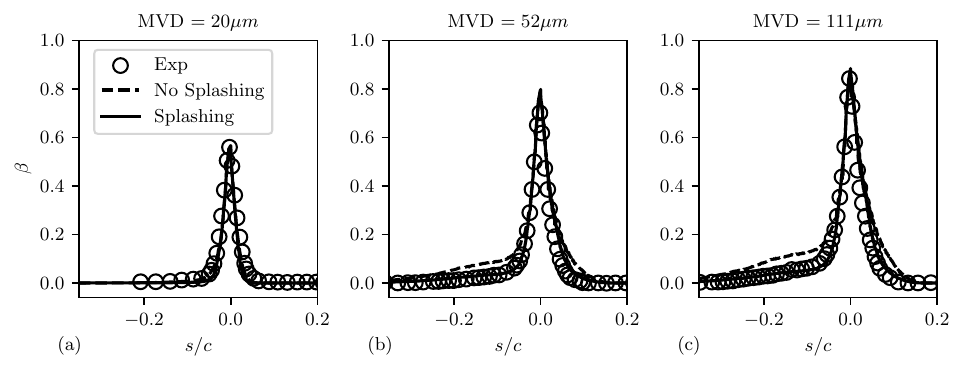}
    \caption{Numerical collection efficiency with and without the splashing model \cite{wrightFurtherRefinementLEWICE2006}, compared to experimental data \cite{papadakisWaterImpingementExperiments2004} for MVDs of (a) 21, (b) 52, and (c) 111 $\upmu$m.}
    \label{fig:SLD_dist}
\end{figure}

The clouds corresponding to MVD of 52 $\upmu$m and 111 $\upmu$m are composed mainly of SLDs, in which the splashing effects are important. To evaluate the effect of splashing, and to assess the model proposed by Wright \cite{wrightFurtherRefinementLEWICE2006} within a LES framework, the collection efficiency for the three particle clouds (MVD $=20\upmu$m, $52\upmu$m, $111\upmu$m) is calculated with and without splashing models. The simulations are performed following the same procedure as described in the previous section, using seven bins to describe the particle cloud. The distributions obtained with the numerical model, superimposed with the experimental results, are presented in Figure \ref{fig:SLD_dist}. For MVD $ = 20\upmu$m, the results are almost identical with and without the splashing model, only showing a small difference on the tail of the distribution on the pressure side. The results of the numerical model match closely the experimental results. For larger values of MVD, when splashing is not considered, the model is capable of representing the collection efficiency near the stagnation point, but overestimates the distribution of $\beta$ on the tails of the distribution, particularly for the pressure side of the airfoil. As mentioned in the previous section, the largest droplets have a bigger influence on the impingement rates away from the stagnation point. Furthermore, away from the stagnation point, droplets impact the airfoil with shallower angles of attack, making the splashing effects more prominent. When accounting for splashing effects, the collection efficiencies calculated with the numerical model follow the experiments closely. These results indicate that the model proposed by Wright \cite{wrightFurtherRefinementLEWICE2006} is suitable to represent splashing for LES with lagrangian particles.

\subsection{Swept Tail NACA 64A008}

The second validation case replicated the experiments by \citet{papadakisExperimentalInvestigationWater2002} for a clean NACA 64A008 swept tail at a 6$^\circ$ angle of attack in droplet clouds with median volume diameters of 21 and 92~$\upmu$m. With a mean aerodynamic chord (MAC) of 36 inches and a freestream velocity of 78.68 m/s, the flow conditions yield a Reynolds number of 5 million and a Mach number of 0.23. The droplet distributions follow a 27-bin characterization, detailed in Table \ref{tab:droplet_dist}. Experimental data from NASA's Glenn Research Center include $C_p$ distributions at midboard and outboard sections and collection efficiency measurements from blotter strips positioned 36 inches from the tunnel floor. The test section dimensions are 9 feet wide by 6 feet tall.

\input{table2}

\subsubsection{Computational setup}

The mesh resolution is 1250 cells per mean aerodynamic chord (MAC$/\Delta_{\min}$), resulting in a total cell count of 20 million control volumes. The computational domain extends 9 MAC streamwise, with cross-sectional dimensions matching the wind tunnel. Slip boundary conditions are applied to the lateral walls, freestream velocity is specified at the inlet, and a nonreflective condition is used at the outlet. Particles are introduced at a rectangular plane 0.5 MAC upstream of the airfoil base with freestream velocity and are removed upon passing the airfoil. Following the results from Section \ref{subsec:NACA23012}, the 27-bin distribution is subsampled to seven representative bins, each simulated separately and then combined on their percentage of LWC. Approximately 60 million particles are injected per convective timescale ($\text{MAC}/U_\infty$) to accelerate the convergence rate.

\subsubsection{Results}
\begin{figure}
    \centering
    \includegraphics{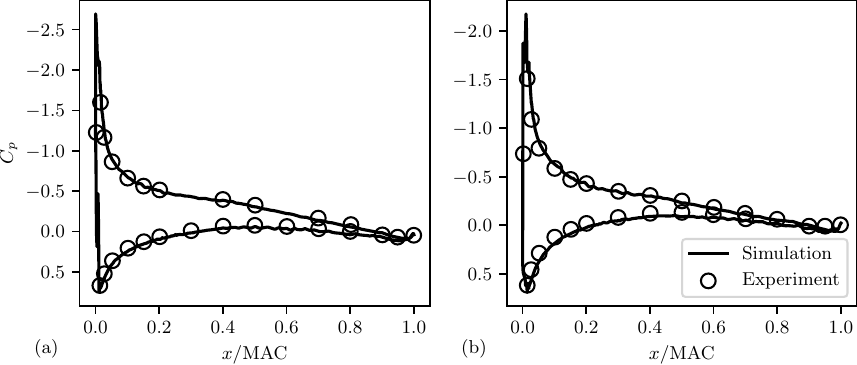}
    \caption{Pressure coefficient ($C_p$) obtained with the numerical model superimposed with the experimental values for the midboard (a) and the outboard (b) sections.}
    \label{fig:CpSwept}
\end{figure}

The simulation was run to a statistically stationary state before introducing droplets into the domain. Statistical stationarity was achieved after 20 convective time units. Figure \ref{fig:CpSwept} presents the $C_p$ distributions for the midboard and outboard sections, demonstrating good agreement between the numerical model and experimental data and validating the aerodynamic part of the calculations.

\begin{figure}
    \centering
    \includegraphics{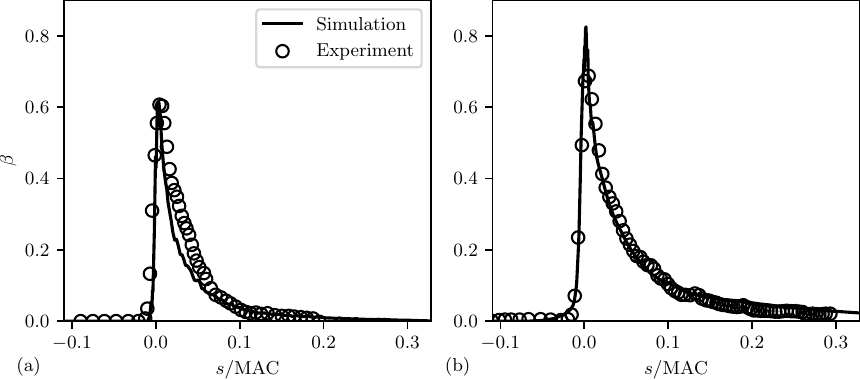}
    \caption{Numerical prediction of collection efficiency ($\beta$) compared to experimental data \cite{papadakisExperimentalInvestigationWater2002} for the (a) 21 $\upmu$m and (b) 92 $\upmu$m MVD clouds on the swept tail.}
    \label{fig:BetaSwept}
\end{figure}

Once the aerodynamic simulation reaches a statistically stationary state, particles are injected into the flow and the simulation is run for another 30 convective time units. Figure \ref{fig:BetaSwept} compares the numerically predicted collection efficiency distributions with the experimental measurements by \citet{papadakisExperimentalInvestigationWater2002} for the two  cloud diameters. For the MVD~21~$\upmu$m case, the numerical model demonstrates good overall agreement with the experimental data, correctly capturing the shape, peak, and impingement limits. A slight underprediction of the  collection efficiency is observed near $x/\text{MAC}=0.03$. This discrepancy is consistent with similar findings from participants of the IPW 1. For the  MVD~=~92~$\upmu$m case, the model's performance is excellent. The simulation results match the experimental data across the entire impingement region, accurately capturing the peak value of $\beta$ and the subsequent distribution. This agreement is notably better than the results presented by most participants from the IPW 1 for this large-droplet condition. While the localized discrepancy near $s$/MAC=0.03 appears to fall outside the 12\% experimental uncertainty band reported by \citet{papadakisExperimentalInvestigationWater2002}, the excellent agreement elsewhere, particularly for the SLD case of MVD~=~92~$\upmu$m, demonstrates the overall high accuracy of the model.

The predictive accuracy demonstrated in these two test cases provides a strong foundation and high confidence in the model's ability to predict droplet impingement. This validation allows us to explore droplet impingement in more complex scenarios for which validation data are not available.

\section{Droplet impingement in laser-scanned airfoils}\label{sec:Roughness}

Following validation on clean geometries, the model is applied to predict droplet impingement on realistic, laser-scanned ice shapes. The geometries selected for this analysis are the rime ice accretions on a NACA 23012 airfoil from the experiments performed by \citet{broerenThreeDimensionalIceAccretionMeasurement2018}. While previous work successfully predicted the aerodynamic degradation for these specific shapes \citep{bornhoftLargeeddySimulationsNACA230122024,bornhoftRoughnessModelingInvestigation2024}, the current investigation focuses on their droplet impingement characteristics, data that were not measured in the original experimental campaign.

To provide a direct, physically consistent evaluation, the simulations are conducted using the exact experimental flow conditions and droplet size distributions that generated these ice shapes. Both accretions were formed at a freestream velocity of $U_\infty = 102.89$ m/s (200 kt), an angle of attack of $2^\circ$, a LWC of 0.4 g/m$^3$, and a static freestream temperature of $-23.1^\circ$C. The droplet cloud is modeled using the experimental 7-bin distribution with a median volume diameter of 30 $\upmu$m. This distribution consists of droplet diameters of 6.9, 9.8, 14.7, 30.3, 60.5, 100.4, and 163.8 $\upmu$m, corresponding to LWC fractions of 5\%, 10\%, 20\%, 30\%, 20\%, 10\%, and 5\%, respectively.

To investigate the temporal evolution of the droplet--surface interaction, two distinct exposure times from this single icing condition are analyzed. The first case, early-stage rime ice (ED1983), corresponds to an exposure time of 1 minute, while the second case, streamwise ice (ED1977), represents the ice shape after 5 minutes of exposure. By fully resolving the scanned roughness elements of these geometries, we can explicitly evaluate how small-scale surface features influence local droplet impingement rates and its distribution.

\subsection{Computational setup}

The computational domain extends seven chord lengths in the streamwise direction and has a spanwise width of $0.25c$.  Similar boundary conditions as used in the previous cases are employed. Table \ref{tab:mesh_roughness} summarizes the computational mesh details and resolution metrics for the two geometries. To accurately capture the small surface features, the early-stage rime case employs a very fine grid comprising 486 million control volumes. The streamwise ice case, which features larger roughness structures, employs 34 million control volumes. For both geometries, the local mesh resolution fully resolves the surface roughness, maintaining $k_{\text{rms}}/\Delta_{\min} > 3$ and $k_{\max}/\Delta_{\min} > 12$, where $k_{\text{rms}}$ is the root-mean-square roughness height, and $k_{\max}$ is the maximum peak-to-trough height. This grid density is considerably finer than that required to achieve accurate aerodynamic load predictions at low angles of attack \cite{bornhoftLargeeddySimulationsNACA230122024}; however, this high resolution was deliberately chosen to ensure that the small-scale roughness features were adequately resolved, which is critical for capturing their effect on the local collection efficiency, $\beta$.

\begin{table}[htbp]
    \centering
    \caption{Computational mesh details and resolution metrics: cell counts $N_{\text{cv}}$; ratio of root-mean-square roughness height ($k_{\text{rms}}$) and maximum peak-to-trough roughness height ($k_{\max}$) to minimum grid length scale ($\Delta_{\min}$).}
    \label{tab:mesh_roughness}
    \begin{tabular}{lccccc}
        \toprule
       Case & Exposure & $N_{\text{cv}}$ [Mcv] & $c / \Delta_{\min}$ & $k_{\text{rms}} / \Delta_{\min}$ & $k_{\max} / \Delta_{\min}$ \\
        \midrule
        Early-stage rime (ED1983) & 1 min & 486 &  14,200 & 3.1 & 12.4 \\
        Streamwise (ED1977) & 5 min &  34 &  3,550 & 3.9 & 15.5 \\
        \bottomrule
    \end{tabular}
\end{table}

For the particle-tracking phase, droplets are introduced at an injection plane located $0.5c$ upstream of the airfoil's leading edge. The rectangular injection area covers the full span of the domain and has a height of $0.22c$. Individual particles are seeded at random locations within this plane. Each particle is tracked until it either intersects the airfoil surface or passes the airfoil trailing edge. As discussed in Section \ref{sec:betaConvergence}, finer computational meshes require significantly longer integration times to achieve statistical convergence. To offset this, the particle injection rates were scaled such that approximately 170 million and 50 million particles were simultaneously tracked within the domain for the early-stage rime ice and streamwise ice cases, respectively.

\subsection{Results}
\begin{figure}
    \centering    
    \includegraphics[width=\textwidth]{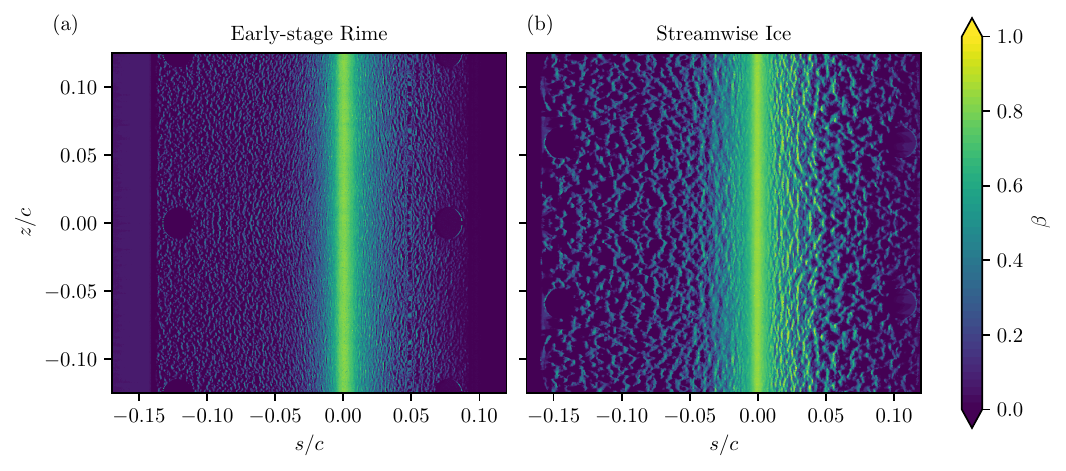}
    \caption{Collection efficiency  distributions over the laser-scanned NACA 23012 ice geometries for (a) the 1-minute early-stage rime accretion and (b) the 5-minute streamwise ice accretion.}
    \label{fig:beta_maps}
\end{figure}

Figure \ref{fig:beta_maps} presents the combined, LWC-weighted collection efficiency maps for the full 30 $\upmu$m MVD droplet cloud over both the early-stage rime (Fig. \ref{fig:beta_maps}-(a)) and streamwise ice (Fig. \ref{fig:beta_maps}-(b)) geometries. The impingement patterns exhibit a highly nonuniform distribution that directly correlates with the underlying surface topology. The 5-minute streamwise ice case, characterized by more pronounced roughness structures, displays a significantly more chaotic and heterogeneous collection efficiency footprint compared to the 1-minute early-rime case. While these maps illustrate the aggregate impingement of the complete droplet cloud, the severity of these localized roughness effects is strongly dependent on droplet inertia.  For small droplets ($<10~\upmu$m), which closely follow the aerodynamic streamlines, impingement is confined primarily near the leading edge where the influence of downstream roughness is limited. As droplet diameter increases, the higher inertia causes the droplets to react more slowly to changes in flow direction, following their original trajectory for longer and widening the impingement limits. For these larger droplets, the effect of surface roughness becomes much more pronounced, with the collection efficiency distribution heavily mirroring the complex texture of the rough ice. A detailed visual breakdown of the collection efficiency maps for individual droplet bin sizes can be found in \citet{zabaletaLargeeddySimulationsDroplet2025}.

\begin{figure}
    \centering
    \includegraphics[width=0.4\textwidth]{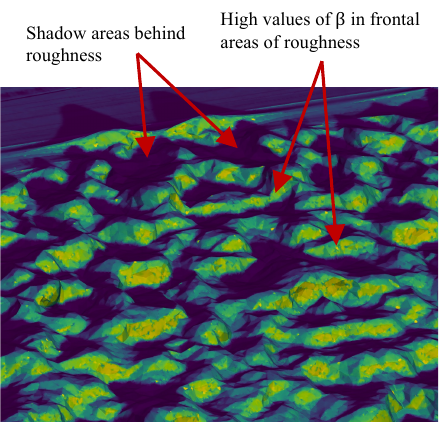} 
    \caption{Magnified three-dimensional view of the local collection efficiency on the streamwise ice roughness elements.}
    \label{fig:beta_closeup}
\end{figure}

Figure \ref{fig:beta_closeup} provides a magnified, three-dimensional view of $\beta$ on the rough surface, revealing that droplet collection is intensely concentrated on the upstream-facing areas of individual roughness elements. Because these protruding features effectively intercept the incoming droplet trajectories, they cast sheltered "shadow" zones immediately downstream where the local collection efficiency drops to near zero. This localized physical interaction is critical to the subsequent ice accretion process. Ultimately, the preferential deposition of incoming water mass on the frontal areas of existing roughness elements creates a self-reinforcing feedback loop, promoting their continued macroscopic growth in the upstream direction.

\subsection{Effect of roughness}

\begin{figure}
    \centering
    \includegraphics[width=0.4\textwidth]{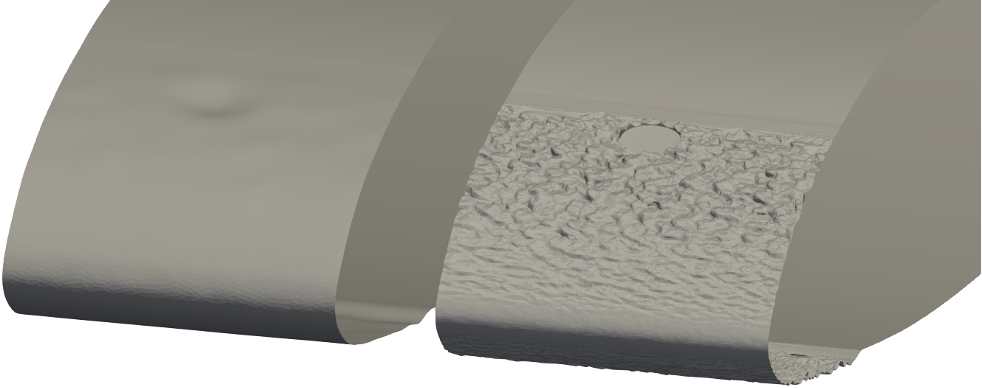} 
    \caption{Comparison of the leading edge of the original laser-scanned geometry (right) and of the equivalent smooth shape (left).}
    \label{fig:smooth_vs_rough}
\end{figure}

To isolate the influence of the measured three-dimensional roughness, simulations are performed on both the laser-scanned geometry and an equivalent smooth body, which was generated by applying a smoothing filtering operator as described by \citet{bellostaAssessingRelevantRoughness2024}. A visual comparison of these two computational configurations is provided in Figure \ref{fig:smooth_vs_rough}. As illustrated, the smooth geometry (left) captures the macroscopic leading-edge deformation characteristic of rime ice accretion, but entirely eliminates the high-frequency, small-scale topological variations present on the laser-scanned surface (right). By comparing the impingement distributions between these two geometries, the specific deviations induced strictly by the physical roughness elements can be decoupled from the global collection efficiency trends.

\begin{figure}
    \centering
    \includegraphics{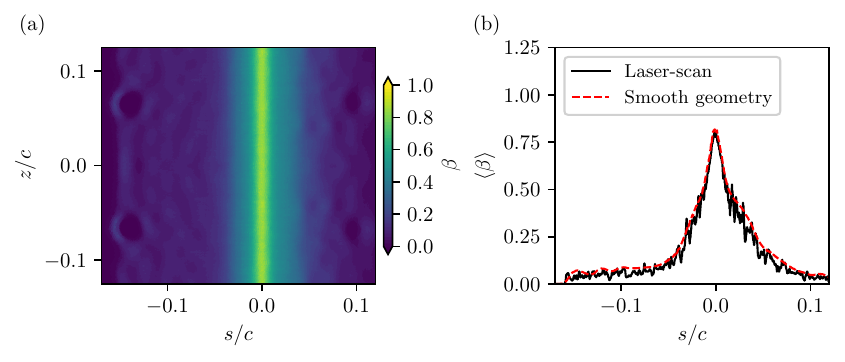}
\caption{(a) Collection efficiency map over the smoothed geometry. (b) Spanwise-averaged collection efficiency ($\langle \beta \rangle$) comparing the laser-scanned rough geometry to the equivalent smooth geometry.}    \label{fig:smooth_vs_rough_beta}
\end{figure}

Figure \ref{fig:smooth_vs_rough_beta} compares the local and spanwise-averaged collection efficiency for both geometries. As shown in the two-dimensional map (Fig.~\ref{fig:smooth_vs_rough_beta}a), the impingement pattern on the smoothed geometry is highly uniform in the spanwise direction. This contrasts the highly localized impingement pattern of the rough surface discussed previously. It should be noted that minor, low-frequency oscillations are visible in the collection efficiency of the smooth shape; these are the result of undulations that remain on the surface after the application of the smoothing filtering operator. Nevertheless, the droplet collection remains smooth, especially compared to the rough case.

Despite these differences in the local droplet redistribution, the spanwise-averaged collection efficiency ($\langle \beta \rangle$) remains remarkably similar between the two shapes. As illustrated in Figure \ref{fig:smooth_vs_rough_beta}b, the spanwise-averaged profile of the laser-scanned geometry closely tracks the macroscopic trend of the smooth shape, although it is characterized by high-frequency spatial oscillations induced by the roughness elements. This indicates that while three-dimensional roughness redistributes the impinging water mass at the roughness scales---concentrating it on forward-facing peaks and sheltering downstream regions---it does not significantly alter the total, macroscopically averaged impingement rate. This observation provides a physical justification for why conventional ice accretion codes, which evaluate droplet impingement over smooth surfaces, can still accurately predict the mean ice shape for rime conditions. While collection efficiency calculated on smooth surfaces neglects local variations in $\beta$, it successfully captures the global mass accumulation required for mean shape prediction.

\section{Roughness evolution}\label{sec:roughness_evolution}

\begin{figure}
    \centering
    \includegraphics{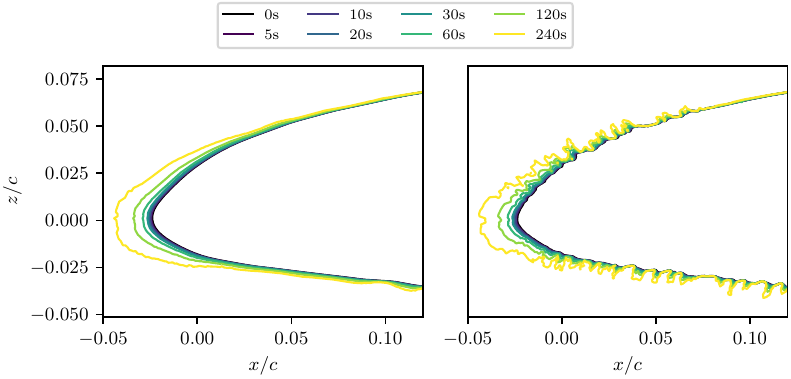}
    \caption{Comparison of the predicted ice growth cross-sections evolving over a 240-second accretion period for the smoothed geometry (left) and the laser-scanned rough geometry (right).}
    \label{fig:iceGrowth}
\end{figure}

The impact of surface roughness on the morphological evolution of the ice is illustrated in Figure \ref{fig:iceGrowth}, which compares the predicted ice growth cross-sections over a 240-second accretion period for both the smoothed and laser-scanned geometries, utilizing the methodology described by \citet{zabaletaMultishotSimulationsRime2025}. It should be noted that this 240-second accretion is intended to be illustrative. While the actual collection efficiency distribution would naturally change as the macroscopic ice geometry evolves, the assumption of a constant collection efficiency during the 240-second accretion process serves to demonstrate how the initial heterogeneity in droplet impingement directly leads to the growth and amplification of surface roughness. These results provide compelling visual evidence for the self-reinforcing nature of surface roughness in ice accretion processes. For the initially smooth surface (left), the simulated ice layer thickens uniformly, closely following the macroscopic contours of the base geometry with minimal development of new high-frequency features. In contrast, the laser-scanned rough surface (right) exhibits a remarkably different and highly irregular growth pattern. The simulation predicts that existing roughness features not only persist but are actively amplified during the accretion process. The initial rough profile becomes progressively more distorted, with pronounced protrusions rapidly growing outward into the flow.

This divergence in growth patterns is a direct consequence of the localized variations in collection efficiency observed previously. On the rough surface, the preferential deposition of droplets on upstream-facing features leads to accelerated growth in these areas, while the sheltered regions experience reduced accretion. 
As a result, the roughness elements grow preferentially in the upstream direction, exaggerating the existing surface irregularities. This mechanism provides a direct physical explanation for the formation and growth of characteristic rime ice structures, and underscores the importance of accurately modeling the local surface geometry. 

\begin{figure}[htbp]
    \centering
    \includegraphics{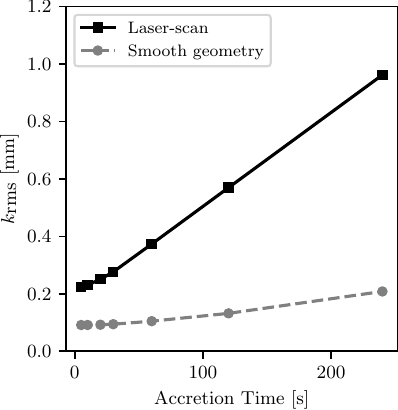}
    \caption{Temporal evolution of the root-mean-square roughness height ($k_{\text{rms}}$) over the 240-second accretion period.}
    \label{fig:krms_evolution}
\end{figure}

To quantify this amplification, the evolution of the root-mean-square roughness height ($k_{\text{rms}}$) over the 240-second accretion window is presented in Figure \ref{fig:krms_evolution}. The $k_{\text{rms}}$ of the laser-scanned geometry experiences rapid, near-linear growth, amplifying from its initial state of approximately 0.2 mm to nearly 1.0 mm by the end of the simulation. Conversely, the roughness metric for the smoothed geometry remains low and exhibits negligible amplification, with its slight increase primarily resulting from the low-frequency oscillations of the original geometry. This divergence in behavior shows that models that evaluate impingement solely over idealized, smooth geometries fundamentally suppress the physical feedback loop required to predict the progressive roughening of the ice surface over time. Furthermore, these findings suggest that even small initial perturbations in surface smoothness could lead to significant roughness development over time.

\section{Conclusions}\label{sec:conclusions}

This study presented a high-fidelity computational framework based on wall-modeled large-eddy simulations (WMLES) and Lagrangian particle tracking to investigate droplet impingement on complex and rough aircraft surfaces. The numerical methodology was first validated against experimental data for a clean NACA 23012 airfoil and a swept NACA 64A008 tail. The framework demonstrated predictable statistical convergence assuming to a spatial Poisson point process, and accurately captured collection efficiency distributions and supercooled large-droplet splashing effects.

Following validation, the model was applied to laser-scanned rime ice geometries to quantify the impact of realistic surface roughness on local impingement. The results revealed that three-dimensional roughness induces a highly nonuniform impingement pattern driven by the roughness topology. Droplet collection efficiency is intensely concentrated on the upstream-facing areas of individual roughness elements, casting sheltered shadow zones immediately downstream. While the spanwise-averaged collection efficiency of the rough surface matches that of an equivalent smooth geometry---providing physical justification for why conventional, smooth-surface ice accretion codes can accurately predict macroscopic mean shapes---the smooth-surface assumption completely filters out the highly localized impingement peaks. 

Finally, ice accretion simulations over a 240-second accretion period provided evidence for the self-reinforcing nature of surface roughness growth. The preferential deposition of water mass on the frontal areas of existing roughness elements actively amplifies surface irregularities, leading to a rapid, near-linear growth in the root-mean-square roughness height ($k_{\text{rms}}$). In contrast, equivalent smooth geometries fundamentally suppress this physical feedback loop, failing to predict the progressive roughening of the ice surface. Ultimately, these findings underscore the critical importance of accurately modeling the local surface geometry during the mass accumulation phase, highlighting that even small initial perturbations can lead to significant roughness development over time.

\section*{Funding Sources}

This investigation was funded by Boeing and NASA's Transformational Tools and Technologies Project. This research used resources of the Oak Ridge Leadership Computing Facility, which is a US Department of Energy, Office of Science, User Facility supported under contract DE-AC0500OR22725.

\bibliography{library_bibtex}

\end{document}

%% file: table1.tex
\begin{table}[htbp]
\centering
\caption{Bin droplet size distribution for 20, 52, and 111 $\upmu$m MVD clouds.}
\label{table:binSizes}
\begin{tabular}{ccccc@{\hspace{2em}}ccccc}
\toprule
\multirow{2}{*}{\textbf{Bin}} & \multirow{2}{*}{\textbf{LWC}} & \multicolumn{3}{c}{\textbf{Droplet sizes by MVD [$\upmu$m]}} & 
\multirow{2}{*}{\textbf{Bin}} & \multirow{2}{*}{\textbf{LWC}} & \multicolumn{3}{c}{\textbf{Droplet sizes by MVD [$\upmu$m]}} \\ 
\cmidrule(lr){3-5} \cmidrule(lr){8-10}
 & & \textbf{20 $\upmu$m} & \textbf{52 $\upmu$m} & \textbf{111 $\upmu$m} & & & \textbf{20 $\upmu$m} & \textbf{52 $\upmu$m} & \textbf{111 $\upmu$m} \\ 
\midrule
1 & 4.75\% & 3.8 & 6.5 & 10.9 & 15 & 4.75\% & 23.6 & 99.8 & 152.8 \\
2 & 4.75\% & 8.4 & 15.2 & 24.5 & 16 & 4.75\% & 24.7 & 115.9 & 165.9 \\
3 & 4.75\% & 10.1 & 18.6 & 29.6 & 17 & 4.75\% & 26.0 & 138.8 & 179.4 \\
4 & 4.75\% & 11.6 & 21.2 & 35.0 & 18 & 4.75\% & 27.5 & 165.0 & 193.7 \\
5 & 4.75\% & 13.0 & 23.6 & 44.7 & 19 & 4.75\% & 29.3 & 185.6 & 207.2 \\
6 & 4.75\% & 14.3 & 25.9 & 58.3 & 20 & 4.75\% & 31.9 & 202.7 & 219.7 \\
7 & 4.75\% & 15.5 & 28.3 & 70.7 & 21 & 1.00\% & 33.8 & 212.4 & 227.4 \\
8 & 4.75\% & 16.7 & 30.9 & 81.3 & 22 & 1.00\% & 34.8 & 215.6 & 230.1 \\
9 & 4.75\% & 17.7 & 34.5 & 91.2 & 23 & 1.00\% & 36.2 & 219.7 & 237.8 \\
10 & 4.75\% & 18.6 & 40.8 & 100.9 & 24 & 0.50\% & 37.5 & 223.6 & 250.5 \\
11 & 4.75\% & 19.5 & 51.4 & 110.6 & 25 & 0.50\% & 38.7 & 226.3 & 264.2 \\
12 & 4.75\% & 20.5 & 63.1 & 119.5 & 26 & 0.50\% & 40.7 & 229.0 & 279.5 \\
13 & 4.75\% & 21.5 & 74.0 & 128.8 & 27 & 0.50\% & 44.4 & 253.9 & 312.6 \\
14 & 4.75\% & 22.5 & 85.7 & 140.1 & & & & & \\ 
\bottomrule
\end{tabular}%
\end{table}

%% file: table2.tex
\begin{table}[htbp]
\centering
\caption{Discretized droplet size distribution for clouds with median volume diameters (MVDs) of 21 $\upmu$m and 92 $\upmu$m. The table shows the liquid water content (LWC) in percent along with the representative droplet diameter for each bin.}
\label{tab:droplet_dist}
\begin{tabular}{cccc@{\hspace{2em}}cccc}
\toprule
\multirow{2}{*}{\textbf{Bin}} & \multirow{2}{*}{\textbf{LWC}} & \multicolumn{2}{c}{\textbf{Droplet sizes by MVD [$\upmu$m]}} & 
\multirow{2}{*}{\textbf{Bin}} & \multirow{2}{*}{\textbf{LWC}} & \multicolumn{2}{c}{\textbf{Droplet sizes by MVD [$\upmu$m]}} \\ 
\cmidrule(lr){3-4} \cmidrule(lr){7-8}
 & & \textbf{21 $\upmu$m} & \textbf{92 $\upmu$m} & & & \textbf{21 $\upmu$m} & \textbf{92 $\upmu$m} \\ 
\midrule
1 & 4.75\% & 4.0 & 10.8 & 15 & 4.75\% & 25.7 & 121.8 \\
2 & 4.75\% & 8.8 & 24.4 & 16 & 4.75\% & 27.1 & 130.6 \\
3 & 4.75\% & 10.3 & 29.8 & 17 & 4.75\% & 28.8 & 140.6 \\
4 & 4.75\% & 11.8 & 35.2 & 18 & 4.75\% & 30.8 & 152.3 \\
5 & 4.75\% & 13.2 & 42.0 & 19 & 4.75\% & 34.4 & 165.6 \\
6 & 4.75\% & 14.7 & 51.4 & 20 & 4.75\% & 47.9 & 187.6 \\
7 & 4.75\% & 16.1 & 61.5 & 21 & 1.00\% & 61.9 & 209.6 \\
8 & 4.75\% & 17.4 & 69.9 & 22 & 1.00\% & 68.7 & 226.1 \\
9 & 4.75\% & 18.5 & 77.5 & 23 & 1.00\% & 76.9 & 247.1 \\
10 & 4.75\% & 19.7 & 84.7 & 24 & 0.50\% & 84.8 & 265.7 \\
11 & 4.75\% & 20.8 & 91.7 & 25 & 0.50\% & 92.4 & 290.3 \\
12 & 4.75\% & 22.0 & 98.8 & 26 & 0.50\% & 103.4 & 339.5 \\
13 & 4.75\% & 23.2 & 106.2 & 27 & 0.50\% & 164.0 & 391.8 \\
14 & 4.75\% & 24.4 & 113.9 & & & & \\ 
\bottomrule
\end{tabular}%

\end{table}

%% file: main.bbl
\begin{thebibliography}{41}
\newcommand{\enquote}[1]{``#1''}
\providecommand{\natexlab}[1]{#1}
\providecommand{\url}[1]{\texttt{#1}}
\providecommand{\urlprefix}{URL }
\expandafter\ifx\csname urlstyle\endcsname\relax
  \providecommand{\doi}[1]{\discretionary{}{}{}https://doi.org/#1}\else
  \providecommand{\doi}[1]{\discretionary{}{}{}\urlstyle{rm}\url{https://doi.org/#1}}\fi

\bibitem[{Gent et~al.(2000)Gent, Dart, and Cansdale}]{gentAircraftIcing2000}
Gent, R.~W., Dart, N.~P., and Cansdale, J.~T., \enquote{Aircraft Icing,} \emph{Philosophical Transactions of the Royal Society of London. Series A: Mathematical, Physical and Engineering Sciences}, Vol. 358, No. 1776, 2000, pp. 2873--2911.
\newblock \doi{10.1098/rsta.2000.0689}.

\bibitem[{Petty and Floyd(2004)}]{pettyStatisticalReviewAviation2004}
Petty, K., and Floyd, C. D.~J., \enquote{Statistical {{Review}} of {{Aviation Airframe Icing}} {{Accidents}} in the {{U}}.{{S}},} \emph{Proceedings of the 11th {{Conference}} on {{Aviation}}, {{Range}}, and {{Aerospace Meteorology}}}, American Meteorology Society, 2004.

\bibitem[{{National Transportation Safety Board}(2023)}]{nationaltransportationsafetyboardNTSBAviationAccidents2023}
{National Transportation Safety Board}, \enquote{{{NTSB}} - {{Aviation Accidents}} - {{Index}} of {{Months}},} https://www.ntsb.gov/safety/Pages/research.aspx, 2023.

\bibitem[{Zabaleta et~al.(2025{\natexlab{a}})Zabaleta, Bornhoft, Jain, Bose, and Moin}]{zabaletaLargeeddySimulationsConjugate2025}
Zabaleta, F., Bornhoft, B., Jain, S.~S., Bose, S.~T., and Moin, P., \enquote{Large-Eddy Simulations of Conjugate Heat Transfer in Boundary Layers over Laser-Scanned Ice Roughness,} \emph{Physical Review Fluids}, Vol.~10, No.~10, 2025{\natexlab{a}}, p. 104603.
\newblock \doi{10.1103/fkg4-y3vf}.

\bibitem[{Moula and Ozcer(2023)}]{moulaIcingSimulationResults2023}
Moula, G., and Ozcer, I., \enquote{Icing {{Simulation Results Using Lagrangian Particle Tracking}} in {{Ansys Fluent Icing}},} \emph{International {{Conference}} on {{Icing}} of {{Aircraft}}, {{Engines}}, and {{Structures}}}, Vienna, Austria, 2023, pp. 2023--01--1478.
\newblock \doi{10.4271/2023-01-1478}.

\bibitem[{Bellosta et~al.(2023)Bellosta, Baldan, Sirianni, and Guardone}]{bellostaLagrangianEulerianAlgorithms2023}
Bellosta, T., Baldan, G., Sirianni, G., and Guardone, A., \enquote{Lagrangian and {{Eulerian}} Algorithms for Water Droplets in In-Flight Ice Accretion,} \emph{Journal of Computational and Applied Mathematics}, Vol. 429, 2023, p. 115230.
\newblock \doi{10.1016/j.cam.2023.115230}.

\bibitem[{Messinger(1953)}]{messingerEquilibriumTemperatureUnheated1953}
Messinger, B.~L., \enquote{Equilibrium {{Temperature}} of an {{Unheated Icing Surface}} as a {{Function}} of {{Air Speed}},} \emph{Journal of the Aeronautical Sciences}, Vol.~20, No.~1, 1953, pp. 29--42.
\newblock \doi{10.2514/8.2520}.

\bibitem[{Myers(2001)}]{myersExtensionMessingerModel2001}
Myers, T.~G., \enquote{Extension to the {{Messinger Model}} for {{Aircraft Icing}},} \emph{AIAA Journal}, Vol.~39, No.~2, 2001, pp. 211--218.
\newblock \doi{10.2514/2.1312}.

\bibitem[{Shad et~al.(2025)Shad, Ahmed, Zgheib, Balachandar, and Sherif}]{shadStokesdependentDropletCollection2025}
Shad, A., Ahmed, H., Zgheib, N., Balachandar, S., and Sherif, S.~A., \enquote{Stokes-Dependent Droplet Collection Efficiency on a {{NACA}} 0012 Airfoil from Droplet-Informed Simulations with Statistical Overloading,} \emph{Philosophical Transactions of the Royal Society A: Mathematical, Physical and Engineering Sciences}, Vol. 383, No. 2301, 2025, p. 20240368.
\newblock \doi{10.1098/rsta.2024.0368}.

\bibitem[{Trontin et~al.(2017)Trontin, Blanchard, Kontogiannis, and Villedieu}]{trontinDescriptionAssessmentNew2017}
Trontin, P., Blanchard, G., Kontogiannis, A., and Villedieu, P., \enquote{Description and Assessment of the New {{ONERA 2D}} Icing Suite {{IGLOO2D}},} \emph{9th {{AIAA Atmospheric}} and {{Space Environments Conference}}}, {American Institute of Aeronautics and Astronautics}, 2017.
\newblock \doi{10.2514/6.2017-3417}.

\bibitem[{Araujo Lima Da~Silva(2025)}]{araujolimadasilvaAdvancingIceAccretionFoamSolver2025}
Araujo Lima Da~Silva, G., \enquote{Advancing the {{iceAccretionFoam Solver}}: {{Glaze Ice Accretion}},} \emph{Philosophical Transactions A}, Vol. 383, No. 2301, 2025.
\newblock \doi{10.1098/rsta.2024.0361open_in_new}.

\bibitem[{Radenac et~al.(2023)Radenac, Blanchard, Renaud, and Duchayne}]{radenacWorkflowPredictorCorrector2023}
Radenac, E., Blanchard, G., Renaud, T., and Duchayne, Q., \enquote{Workflow for Predictor--Corrector Simulations of in-Flight Ice Accretion, with Applications on Swept Wings,} \emph{Engineering with Computers}, 2023.
\newblock \doi{10.1007/s00366-023-01910-y}.

\bibitem[{Guardone et~al.(2025)Guardone, Bellosta, Donizetti, and Gallia}]{guardoneAircraftIcingModeling2025}
Guardone, A., Bellosta, T., Donizetti, A., and Gallia, M., \enquote{Aircraft {{Icing}}: {{Modeling}} and {{Simulation}},} \emph{Annual Review of Fluid Mechanics}, 2025.
\newblock \doi{10.1146/annurev-fluid-112723-062537}.

\bibitem[{Laurendeau et~al.(2022)Laurendeau, {Bourgault-Cote}, Ozcer, Hann, Radenac, and Pueyo}]{laurendeauSummary1stAIAA2022}
Laurendeau, E., {Bourgault-Cote}, S., Ozcer, I.~A., Hann, R., Radenac, E., and Pueyo, A., \enquote{Summary from the 1st {{AIAA Ice Prediction Workshop}},} \emph{{{AIAA AVIATION}} 2022 {{Forum}}}, {American Institute of Aeronautics and Astronautics}, 2022.
\newblock \doi{10.2514/6.2022-3398}.

\bibitem[{Ignatowicz et~al.(2023)Ignatowicz, Morency, and Beaugendre}]{ignatowiczSurfaceRoughnessRANS2023}
Ignatowicz, K., Morency, F., and Beaugendre, H., \enquote{Surface {{Roughness}} in {{RANS Applied}} to {{Aircraft Ice Accretion Simulation}}: {{A Review}},} \emph{Fluids}, Vol.~8, No.~10, 2023, p. 278.
\newblock \doi{10.3390/fluids8100278}.

\bibitem[{Ozcer et~al.(2011)Ozcer, Baruzzi, Reid, Habashi, Fossati, and Croce}]{ozcerFENSAPICENumericalPrediction2011}
Ozcer, I.~A., Baruzzi, G.~S., Reid, T., Habashi, W.~G., Fossati, M., and Croce, G., \enquote{{{FENSAP-ICE}}: {{Numerical Prediction}} of {{Ice Roughness Evolution}}, and Its {{Effects}} on {{Ice Shapes}},} \emph{{{SAE}} 2011 {{International Conference}} on {{Aircraft}} and {{Engine Icing}} and {{Ground Deicing}}}, 2011, pp. 2011--38--0024.
\newblock \doi{10.4271/2011-38-0024}.

\bibitem[{Bellosta et~al.(2024)Bellosta, Donizetti, Zabaleta, Bornhoft, Jain, Bose, and Guardone}]{bellostaAssessingRelevantRoughness2024}
Bellosta, T., Donizetti, A., Zabaleta, F., Bornhoft, B., Jain, S.~S., Bose, S.~T., and Guardone, A., \enquote{Assessing Relevant Roughness Scales for Accurate Prediction of Iced Airfoil Aerodynamics,} \emph{Proceedings of the 2024 {{Summer Program}}}, Center for Turbulence Research, Stanford University, 2024, pp. 427--437.

\bibitem[{Bornhoft et~al.(2025)Bornhoft, Moin, Jain, and Bose}]{bornhoftUseArtificialIce2025}
Bornhoft, B., Moin, P., Jain, S.~S., and Bose, S.~T., \enquote{On the {{Use}} of {{Artificial Ice Shapes}} for {{Large-Eddy Simulations}} in {{Aircraft Icing}},} \emph{Journal of Aircraft}, 2025, pp. 1--18.
\newblock \doi{10.2514/1.C038146}.

\bibitem[{Freschi et~al.(2025)Freschi, Donizetti, Bellosta, and Guardone}]{freschiTwoDimensionalMultiStepStochastic2025}
Freschi, M., Donizetti, A., Bellosta, T., and Guardone, A., \enquote{A {{Two-Dimensional Multi-Step Stochastic Approach}} for {{Straight Wing Ice Accretion Analyses}},} \emph{{{AIAA AVIATION FORUM AND ASCEND}} 2025}, {American Institute of Aeronautics and Astronautics}, Las Vegas, Nevada, 2025.
\newblock \doi{10.2514/6.2025-3515}.

\bibitem[{Zabaleta et~al.(2025{\natexlab{b}})Zabaleta, Yu, and Moin}]{zabaletaMultishotSimulationsRime2025}
Zabaleta, F., Yu, H., and Moin, P., \enquote{Multi-Shot Simulations of Rime Ice Accretion,} \emph{Annual {{Research Briefs}}}, Center for Turbulence Research, Stanford University, 2025{\natexlab{b}}.

\bibitem[{Br{\`e}s et~al.(2018)Br{\`e}s, Bose, Emory, Ham, Schmidt, Rigas, and Colonius}]{bresLargeeddySimulationsCoannular2018}
Br{\`e}s, G.~A., Bose, S.~T., Emory, M., Ham, F.~E., Schmidt, O.~T., Rigas, G., and Colonius, T., \enquote{Large-Eddy Simulations of Co-Annular Turbulent Jet Using a {{Voronoi-based}} Mesh Generation Framework,} \emph{2018 {{AIAA}}/{{CEAS Aeroacoustics Conference}}}, {American Institute of Aeronautics and Astronautics}, Atlanta, Georgia, 2018.
\newblock \doi{10.2514/6.2018-3302}.

\bibitem[{Germano et~al.(1991)Germano, Piomelli, Moin, and Cabot}]{germanoDynamicSubgridScale1991}
Germano, M., Piomelli, U., Moin, P., and Cabot, W.~H., \enquote{A Dynamic Subgrid-scale Eddy Viscosity Model,} \emph{Physics of Fluids A: Fluid Dynamics}, Vol.~3, No.~7, 1991, pp. 1760--1765.
\newblock \doi{10.1063/1.857955}.

\bibitem[{Moin et~al.(1991)Moin, Squires, Cabot, and Lee}]{moinDynamicSubgridScale1991}
Moin, P., Squires, K., Cabot, W., and Lee, S., \enquote{A Dynamic Subgrid-scale Model for Compressible Turbulence and Scalar Transport,} \emph{Physics of Fluids A: Fluid Dynamics}, Vol.~3, No.~11, 1991, pp. 2746--2757.
\newblock \doi{10.1063/1.858164}.

\bibitem[{Honein and Moin(2004)}]{honeinHigherEntropyConservation2004}
Honein, A.~E., and Moin, P., \enquote{Higher Entropy Conservation and Numerical Stability of Compressible Turbulence Simulations,} \emph{Journal of Computational Physics}, Vol. 201, No.~2, 2004, pp. 531--545.
\newblock \doi{10.1016/j.jcp.2004.06.006}.

\bibitem[{Chandrashekar(2013)}]{chandrashekarKineticEnergyPreserving2013}
Chandrashekar, P., \enquote{Kinetic {{Energy Preserving}} and {{Entropy Stable Finite Volume Schemes}} for {{Compressible Euler}} and {{Navier-Stokes Equations}},} \emph{Communications in Computational Physics}, Vol.~14, No.~5, 2013, pp. 1252--1286.
\newblock \doi{10.4208/cicp.170712.010313a}.

\bibitem[{Lehmkuhl et~al.(2018)Lehmkuhl, Park, Bose, and Moin}]{lehmkuhlLargeeddySimulationPractical2018}
Lehmkuhl, O., Park, G.~I., Bose, S.~T., and Moin, P., \enquote{Large-Eddy Simulation of Practical Aeronautical Flows at Stall Conditions,} \emph{{{CTR Proceedings}} of {{Summer Program}}}, 2018, pp. 87--96.

\bibitem[{Elghobashi(1994)}]{Elghobashi1994}
Elghobashi, S., \enquote{On Predicting Particle-Laden Turbulent Flows,} \emph{Applied Scientific Research}, Vol.~52, No.~4, 1994, pp. 309--329.
\newblock \doi{10.1007/BF00936835}.

\bibitem[{Schiller and Naumann(1935)}]{schillerDragCoefficientCorrelation1935}
Schiller, L., and Naumann, A., \enquote{A {{Drag Coefficient Correlation}},} \emph{Zeitschrift des Vereins Deutscher Ingenieure}, Vol.~77, 1935, pp. 318--320.

\bibitem[{M{\"o}ller and Trumbore(2005)}]{mollerFastMinimumStorage2005}
M{\"o}ller, T., and Trumbore, B., \enquote{Fast, Minimum Storage Ray/Triangle Intersection,} \emph{{{ACM SIGGRAPH}} 2005 {{Courses}} on - {{SIGGRAPH}} '05}, ACM Press, Los Angeles, California, 2005, p.~7.
\newblock \doi{10.1145/1198555.1198746}.

\bibitem[{Wright(2006)}]{wrightFurtherRefinementLEWICE2006}
Wright, W., \enquote{Further {{Refinement}} of the {{LEWICE SLD Model}},} \emph{44th {{AIAA Aerospace Sciences Meeting}} and {{Exhibit}}}, Aerospace {{Sciences Meetings}}, {American Institute of Aeronautics and Astronautics}, 2006.
\newblock \doi{10.2514/6.2006-464}.

\bibitem[{Wright et~al.(2008)Wright, Potapczuk, and Levinson}]{wrightComparisonLEWICEGlennICE2008a}
Wright, W., Potapczuk, M., and Levinson, L., \enquote{Comparison of {{LEWICE}} and {{GlennICE}} in the {{SLD Regime}},} \emph{46th {{AIAA Aerospace Sciences Meeting}} and {{Exhibit}}}, {American Institute of Aeronautics and Astronautics}, Reno, Nevada, 2008.
\newblock \doi{10.2514/6.2008-439}.

\bibitem[{Mundo et~al.(1995)Mundo, Sommerfeld, and Tropea}]{mundoDropletwallCollisionsExperimental1995}
Mundo, {\relax Chr}., Sommerfeld, M., and Tropea, C., \enquote{Droplet-Wall Collisions: {{Experimental}} Studies of the Deformation and Breakup Process,} \emph{International Journal of Multiphase Flow}, Vol.~21, No.~2, 1995, pp. 151--173.
\newblock \doi{10.1016/0301-9322(94)00069-V}.

\bibitem[{Drew and Passman(1998)}]{drew1998theory}
Drew, D.~A., and Passman, S.~L., \emph{Theory of {{Multicomponent Fluids}}}, Springer New York, 1998.

\bibitem[{Papadakis et~al.(2004)Papadakis, Rachman, Wong, Yeong, Hung, and Bidwell}]{papadakisWaterImpingementExperiments2004}
Papadakis, M., Rachman, A., Wong, S.-C., Yeong, H.-W., Hung, K., and Bidwell, C., \enquote{Water {{Impingement Experiments}} on a {{NACA}} 23012 {{Airfoil}} with {{Simulated Glaze Ice Shapes}},} {American Institute of Aeronautics and Astronautics}, 2004.
\newblock \doi{10.2514/6.2004-565}.

\bibitem[{Air(2014)}]{AirplaneEngineCertification2014}
\enquote{Airplane and {{Engine Certification Requirements}} in {{Supercooled Large Drop}}, {{Mixed Phase}}, and {{Ice Crystal Icing Conditions}},} Rule 2014-25789, Federal Aviation Administration, Nov. 2014.

\bibitem[{Zabaleta et~al.(2024)Zabaleta, Jain, Bornhoft, Bose, and Moin}]{zabaletaLargeEddySimulationSupercooled2024}
Zabaleta, F., Jain, S.~S., Bornhoft, B.~J., Bose, S., and Moin, P., \enquote{Large-{{Eddy Simulation}} of {{Supercooled Large Droplets Impingement Using}} a {{Lagrangian Particle Approach}},} \emph{{{AIAA AVIATION FORUM AND ASCEND}} 2024}, {American Institute of Aeronautics and Astronautics}, Las Vegas, Nevada, 2024.
\newblock \doi{10.2514/6.2024-4162}.

\bibitem[{Papadakis et~al.(2002)Papadakis, Hung, Vu, Yeong, Bidwell, Breer, and Bencic}]{papadakisExperimentalInvestigationWater2002}
Papadakis, M., Hung, K.~E., Vu, G.~T., Yeong, H.~W., Bidwell, C.~S., Breer, M.~D., and Bencic, T.~J., \enquote{Experimental {{Investigation}} of {{Water Droplet Impingement}} on {{Airfoils}}, {{Finite Wings}}, and an {{S-duct Engine Inlet}},} Technical {{Memorandum}} 20020090796, 2002.

\bibitem[{Broeren et~al.(2018)Broeren, Addy, Lee, Monastero, and McClain}]{broerenThreeDimensionalIceAccretionMeasurement2018}
Broeren, A.~P., Addy, H.~E., Lee, S., Monastero, M.~C., and McClain, S.~T., \enquote{Three-{{Dimensional Ice-Accretion Measurement Methodology}} for {{Experimental Aerodynamic Simulation}},} \emph{Journal of Aircraft}, Vol.~55, No.~2, 2018, pp. 817--828.
\newblock \doi{10.2514/1.C034580}.

\bibitem[{Bornhoft et~al.(2024{\natexlab{a}})Bornhoft, Jain, Goc, Bose, and Moin}]{bornhoftLargeeddySimulationsNACA230122024}
Bornhoft, B., Jain, S.~S., Goc, K., Bose, S.~T., and Moin, P., \enquote{Large-Eddy Simulations of the {{NACA23012}} Airfoil with Laser-Scanned Ice Shapes,} \emph{Aerospace Science and Technology}, Vol. 146, 2024{\natexlab{a}}, p. 108957.
\newblock \doi{10.1016/j.ast.2024.108957}.

\bibitem[{Bornhoft et~al.(2024{\natexlab{b}})Bornhoft, Jain, Bose, and Moin}]{bornhoftRoughnessModelingInvestigation2024}
Bornhoft, B.~J., Jain, S.~S., Bose, S., and Moin, P., \enquote{Roughness {{Modeling Investigation}} in {{Large-Eddy Simulations}} of a {{NACA23012 Airfoil Under Rime Ice Conditions}},} \emph{{{AIAA AVIATION FORUM AND ASCEND}} 2024}, {American Institute of Aeronautics and Astronautics}, Las Vegas, Nevada, 2024{\natexlab{b}}.
\newblock \doi{10.2514/6.2024-4001}.

\bibitem[{Zabaleta and Moin(2025)}]{zabaletaLargeeddySimulationsDroplet2025}
Zabaleta, F., and Moin, P., \enquote{Large-Eddy Simulations of Droplet Impingement on Iced Airfoils,} \emph{Annual {{Research Briefs}}}, Center for Turbulence Research, Stanford University, 2025.

\end{thebibliography}
